\documentclass[11pt,onecolumn,amssymb,nofootinbib]{revtex4}
\usepackage{amsmath, amsthm, amscd, amssymb}
\usepackage{graphicx, braket}
\usepackage{bm}
\usepackage{bbm}
\usepackage{epstopdf}
\usepackage{latexsym}
\usepackage[format=hang,font=small,labelfont=bf]{caption}
\usepackage[colorlinks=true,citecolor=red,linkcolor=blue,urlcolor=blue]{hyperref}
\newcommand\myoplus{{\stackrel{\kappa}{\oplus}}}
\newcommand\myln{\ln_\kappa}
\newcommand\mye{e_\kappa}
\DeclareMathOperator{\arcsinh}{arcsinh}
    
\begin{document}
\title{Relativistic Correction to Black Hole Entropy}
\author{Naman Kumar}
\email{namankumar5954@gmail.com}
\affiliation{School of Basic Sciences, Indian Institute of Technology Bhubaneswar, Odisha, 752050}

\begin{abstract}
In this paper, we study the relativistic correction to Bekenstein-Hawking entropy in the canonical ensemble and isothermal-isobaric ensemble and apply it to the cases of non-rotating BTZ and AdS-Schwarzschild black holes. This is realized by generalizing the equations obtained using Boltzmann-Gibbs(BG) statistics with its relativistic generalization, Kaniadakis statistics, or $\kappa$-statistics. The relativistic corrections are found to be logarithmic in nature and it is observed that their effect becomes appreciable in the high-temperature limit suggesting that the entropy corrections must include these relativistically corrected terms while taking the aforementioned limit. The non-relativistic corrections are recovered in the $\kappa\to0$ limit.
\end{abstract}
\maketitle
\tableofcontents
\newpage
\section{Introduction}
\numberwithin{equation}{section}

It is known that for large black holes, the Bekenstein-Hawking entropy gets logarithmic corrections\cite{solodukhin1995conical,fursaev1995temperature,mann1996conical,carlip2000logarithmic,govindarajan2001logarithmic}
and have been evaluated for extremal and non-extremal black holes, using, for example, Euclidean quantum gravity methods\cite{banerjee2011logarithmic,sen2012logarithmic,mandal2010black,fursaev1996one} and the saddle point method\cite{kaul2000logarithmic,das2002general,mukherji2002logarithmic}. Black hole thermodynamics have been investigated for the case of canonical ensemble\cite{das2002general} as well as the isothermal-isobaric ensemble\cite{ghosh2022novel} in the extended black hole thermodynamics formalism\cite{kastor2009enthalpy} wherein the concept of pressure arises due to a dynamical cosmological constant. This is readily achieved in AdS spacetime where pressure $P$ and cosmological constant $\Lambda$ are related as
\begin{equation}
    P=-\frac{\Lambda}{8\pi}
\end{equation}
The general form of corrections to the entropy in these ensembles has been found as
\begin{equation}
    \mathcal{S}=S_0-k\ln S_0
\end{equation}
where the constant $k$ depends on the choice of statistical ensemble and the black hole under consideration. All of these analysis were based on the Boltzmann-Gibbs(BG) statistics. A nice discussion on various statistical ensembles in BG statistics and their application for a variety of black holes can be found in\cite{ghosh2023statistical}. However, it is known that BG statistics has severe restrictions in a gravitational setting\cite{tsallis1995some} when applied to a thermally fluctuating system with a long-range force\cite{luciano2022gravity}. 
Therefore, it becomes pertinent to study the entropy corrections to a black hole using some generalized statistics of which BG statistics is a special case. The basic recipe for calculating the entropy corrections using statistical ensembles is as follows: First, we write the partition function as the Laplace transform of the density of states. We then take the Laplace inverse to find the density of states. Taking the natural log of the density of states gives the corrected microcanonical entropy. However, this formalism is not consistent with relativity and therefore does not take into account the relativistic correction that might arise in a thermally fluctuating and long-interacting system such as a black hole. To circumvent these subtle issues we use generalized statistics called Kaniadakis statistics\cite{kaniadakis2001non,kaniadakis2002statistical,kaniadakis2005statistical} which is a relativistic generalization of BG statistics where the integrals and exponentials are replaced by their relativistic counterpart and reduces to BG statistics as a special case (see Appendix A for a brief review). Moreover, Kaniadakis statistics has been recently used to describe modified FLRW cosmology where the entropy of the apparent horizon is taken to be of the form of Kaniadakis entropy\cite{sheykhi2023corrections,lymperis2021modified}. Black hole thermodynamics has also been studied using dual Kaniadakis entropy formalism in\cite{abreu2021black}. 
Therefore, by studying $\kappa$-statistics, we can find potential relativistic corrections to Bekenstein-Hawking entropy not considered before. It is observed that these relativistic corrections become important in the high-temperature limit.\\
The paper is organized as follows: In section II, we calculate the corrected microcanonical entropy for the cases of a canonical ensemble as well as the isothermal-isobaric ensemble. We then apply these corrected microcanonical entropy equations to a class of black holes such as non-rotating BTZ and AdS-Schwarzschild black holes in section III. We conclude the paper with a short discussion in section IV. Two appendices, A and B have also been added to make the presentation self-contained and complete. 

\section{Calculation of Microcanonical Entropy}
In this section, we calculate the microcanonical entropy for the cases of canonical ensemble and \\isothermal-isobaric ensemble using $\kappa$-statistics. Let us first start with the canonical ensemble (one variable case) and then move to the isothermal-isobaric ensemble (two variables case) to explicitly generalize the procedure.
\numberwithin{equation}{section}
\subsection{Canonical Ensemble}
We start by establishing a relation between $\kappa$-entropy ($S_\kappa$) and $\kappa$-deformed partition function ($\mathcal{Z}_\kappa$) in the canonical ensemble. First, the probability distribution in the ordinary statistics is given by
\begin{equation}
    \rho_i=\frac{e^{-\beta E_i}}{\mathcal{Z}}
\end{equation}
where $\beta=1/k_BT$. Analogously, in the $\kappa$-deformed statistics, we can write this in terms of $\kappa$-deformed exponential and partition function as\cite{kaniadakis2001non}
\begin{equation}
      \rho^i_\kappa=\frac{e_\kappa({-\beta E_i})}{\mathcal{Z}_\kappa}\label{k_partition}
\end{equation}
This distribution called the $\kappa$-distribution has been obtained using the maximum entropy principle\cite{kaniadakis2001non}.
The $\kappa$-deformed entropy is given by
\begin{equation}
   S_{\kappa}=-\sum_i\rho^i_\kappa\ln_{\kappa}\rho^i_\kappa\label{kentropy}
\end{equation}
Using this definition of entropy, we obtain\footnote{See Appendix B for a derivation.}\textsuperscript{,}\footnote{We set $k_B=1$ in this paper.}
\begin{equation}
S_\kappa=\ln_{\kappa}\mathcal{Z}_\kappa\hspace{0.5mm}{\myoplus}\hspace{0.5mm}\beta U\label{kentropy2}
\end{equation}
In the limit $\kappa\to0$ we recover the ordinary statistics as
\begin{equation}
    S=\ln\mathcal{Z}+\beta U
\end{equation}
The $\kappa$-partition function is given as the $\kappa$-Laplace transform of the density of states as
\begin{equation}
    \mathcal{Z}_\kappa(\beta)=\int_0^\infty \rho(E)e_\kappa(-\beta E) dE
\end{equation}
To write it in the form of inverse $\kappa$-Laplace transform (Eq.(\ref{lapinv})), we introduce a change of variable as $\epsilon=\beta E$ and introduce a dummy parameter $\zeta$. Then, the $\kappa$-partition function is given as the $\kappa$-Laplace transform of the density of states as
\begin{equation}
    \mathcal{Z}_\kappa(\beta,\zeta)=\int_0^\infty \rho(\epsilon/\beta)[e_\kappa(-\epsilon)]^\zeta d\epsilon\bigg|_{\zeta=1}
\end{equation}
Inverting the equation, we obtain the density of states as
\begin{equation}
    \rho(E)=\frac{1}{2\pi i}\int_{c-i\infty}^{c+i\infty} \frac{\mathcal{Z}_\kappa(\beta,\zeta)[e_\kappa(\epsilon)]^\zeta}{\sqrt{1+\kappa^2\epsilon^2}} d\zeta\bigg|_{\epsilon=\beta E}
\end{equation}
Using Eq.(\ref{kentropy2}), we can write this as
\begin{equation}
    \rho(E)=\frac{1}{2\pi i}\int_{c-i\infty}^{c+i\infty}\frac{e_{\kappa}(S_{\kappa})}{\sqrt{1+\kappa^2\epsilon^2}} d\zeta=\frac{1}{2\pi i}\int_{c-i\infty}^{c+i\infty}\frac{e^S}{\sqrt{1+\kappa^2\epsilon^2}} d\zeta\bigg|_{\epsilon=\beta E}\label{dos}
\end{equation}
where $e_\kappa(S_\kappa)=\mathcal{Z}_\kappa(\beta) e_\kappa(\epsilon)=\mathcal{Z}_\kappa(\beta,\zeta)[e_\kappa(\epsilon)]^\zeta\big|_{\zeta=1}$. We have written the relation $e_\kappa S_\kappa=e^S$ using the definition of $e_\kappa$ as given in Eq.(\ref{ekappa}) and using $S_\kappa=\frac{1}{\kappa}(\sinh\kappa S)$. The latter expression is obtained following\cite{moradpour2020generalized} where we have replaced $S_{BH}$ by the entropy function $S=S(\beta)$ such that $S(\beta)=\ln W$ since we are now also considering fluctuations in the inverse temperature $\beta$. We now expand the entropy function about its equilibrium value $S_0$ and solve Eq.(\ref{dos}) using the steepest descent method, the result is obtained as
\begin{equation}
    \rho(E)=\frac{e^{S_0}}{\sqrt{2\pi S''_0}}\frac{1}{\sqrt{1+\kappa^2\beta_0^2E^2}}
\end{equation}
where $S_0$ is the Bekenstein-Hawking entropy, $\beta_0=1/T_H$ with $T_H$ being Hawking temperature, $E=M$ with $M$ being the mass of the black hole under consideration and $S_0''=\big(\frac{\partial^2S(\beta)}{\partial\beta^2}\big)_{\beta=\beta_0}$. The multiplicative factor $1/\sqrt{1+\kappa^2\beta_0^2E^2}$ can be understood as a Lorentz-like factor for the canonical ensemble. The logarithm of the density of states gives the microcanonical entropy as
\begin{equation}
    \mathcal{S}=S_0-\frac{1}{2}\ln S_0''-\frac{1}{2}\ln(1+\kappa^2\beta_0^2E^2)\label{relmc}
\end{equation}
The third term in Eq.(\ref{relmc}) is the relativistically corrected term. In the limit $\kappa\to0$ we recover the non-relativistic correction as given in\cite{das2002general,ghosh2022novel}. Also, it is straightforward to see that
\begin{equation}
    S''_0=T^2C
\end{equation}
where $C=(\partial E/\partial T)_{T_0}$ is the specific heat. We, therefore, obtain the final expression for the corrected microcanonical entropy in $\kappa$-deformed statistics as
\begin{equation}
    S=S_0-\frac{1}{2}\ln T^2C-\frac{1}{2}\ln(1+\kappa^2\beta_0^2E^2)
\end{equation}
It must be noted that for the above expression to be meaningful $C>0$ which is related to the stability of black holes.
\subsection{Isobaric-Isothermal Ensemble}
We now consider the entropy correction in $\kappa$-deformed statistics due to fluctuations in two variables. For this we consider a $(NPT)$-ensemble, the partition function is given as
\begin{equation}
    \Delta(\beta,\beta P)=C\int_{0}^{\infty}\int_0^\infty \rho(E,V) e^{-\beta(E+PV)} dE dV
\end{equation}
where $C$ is a constant of appropriate dimension to make $\Delta(\beta,\beta P)$ dimensionless and as we will see, the value of $C$ is irrelevant to our calculation.
Equivalently in the $\kappa$-deformed statistics, this can be written as\footnote{Analogous to the single Laplace transform, this is the $\kappa$-deformed generalization of the double Laplace transform.}
\begin{equation}
\Delta_\kappa(\beta,\beta P)=C\int_{0}^{\infty}\int_0^\infty \rho(E,V) [e_\kappa(-\beta E)] [e_\kappa(-\beta PV)] dE dV
\end{equation}
Let us introduce a change of variables with $\epsilon=\beta E$ and $\gamma=\beta PV$ and introduce two dummy parameters $\eta$ and $\zeta$ such that the inverse $\kappa$-deformed Laplace integral is well-defined in the form of Eq.(\ref{lapinv}) as follows
\begin{equation}
 \Delta_\kappa(\beta,\beta P,\eta,\zeta)=C\int_{0}^{\infty}\int_0^\infty \rho(\epsilon/\beta,\gamma/\beta P) [e_\kappa(-\epsilon)]^\eta[e_\kappa(-\gamma)]^\zeta\ d\epsilon d\gamma\bigg|_{\eta=\zeta=1}
\end{equation}
Inverting this equation, we get
\begin{equation}
    \rho(E,V)=\frac{C^{-1}}{(2\pi i)^2}\int_{c-i\infty}^{c+i\infty}\int_{d-i\infty}^{d+\infty} \frac{e_{\kappa} S_{\kappa}}{\sqrt{1+(\kappa\epsilon)^2}\sqrt{1+(\kappa\gamma)^2}}\ d\eta d\zeta\bigg|_{\epsilon=\beta E,\gamma=\beta PV}
\end{equation}
where $e_\kappa(S_\kappa)=\Delta_\kappa(\beta,\beta P)e_\kappa(\beta E)e_\kappa(\beta PV)=\Delta_\kappa(\beta,\beta P,\eta,\zeta)[e_\kappa(\epsilon)]^\eta[e_\kappa(\gamma)]^\zeta\big|_{\eta=\zeta=1}$. The above equation is
equivalent to
\begin{equation}
    \rho(E,V)=\frac{C^{-1}}{(2\pi i)^2}\int_{c-i\infty}^{c+i\infty}\int_{d-i\infty}^{d+\infty} \frac{e^S}{\sqrt{1+(\kappa\epsilon)^2}\sqrt{1+(\kappa\gamma)^2}} d\eta d\zeta \bigg|_{\epsilon=\beta E,\gamma=\beta PV}
\end{equation}
Solving this, we get
\begin{equation}
    \rho(E,V)=\frac{C^{-1}e^{S_0}}{(2\pi)\sqrt{D}\sqrt{1+(\kappa\beta_0 E)^2}\sqrt{1+(\kappa \beta_0 PV)^2}}
\end{equation}
with $D=\partial_\beta^2S_0\partial_{\beta P}^2S_0-(\partial_\beta\partial_{\beta P}S_0)^2$. Therefore, the relativistically corrected microcanonical entropy is obtained by taking the log of the density of states as
\begin{equation}
    \mathcal{S}=\ln\rho(E,V)=S_0-\frac{1}{2}\ln D-\frac{1}{2}\ln(1+(\kappa\beta_0E)^2)-\frac{1}{2}\ln(1+(\kappa\beta_0PV)^2)\label{nptmc}
\end{equation}
The multiplicative factors $\frac{1}{\sqrt{1+(\kappa\beta_0E)^2}}$ and $\frac{1}{\sqrt{1+(\kappa\beta_0PV)^2}}$ can again be understood as the "Lorentz factors" for this particular statistical ensemble. In the limit $\kappa\to0$, we obtain the non-relativistic correction. Also, from Eq.(\ref{nptmc}), it is to be noted that the relativistic corrections become appreciable only in the high-temperature limit.
\section{Application to Black Holes}
In this section, we apply Eq.(\ref{relmc}) and Eq.(\ref{nptmc}) to non-rotating BTZ and AdS-Schwarzschild black holes. Black holes in AdS spacetime have a natural meaning of pressure as a cosmological constant and BTZ black holes are defined with a negative cosmological constant acting as pressure. Therefore, these black holes can be analyzed in both canonical and isothermal-isobaric ensembles and are relevant to our discussion. These black holes have been extensively discussed in various settings in the BG statistics in\cite{ghosh2023statistical}. It is observed that in the high-temperature limit, the relativistic corrections become appreciable. So, we mostly focus on this limit. In the limit $\kappa\to0$, we recover the non-relativistic correction.
\subsection{BTZ Black Hole}
For a non-rotating BTZ black hole in $3D$, the metric is given as
\begin{equation}
    ds^2=-\bigg(\frac{r^2}{l^2}-8G_3M\bigg)dt^2+\bigg(\frac{r^2}{l^2}-8G_3M\bigg)^{-1}dr^2+r^2d\theta^2
\end{equation}
where $G_3$ is the three-dimensional Newton's constant and $l$ is related to the cosmological constant as $\Lambda=-1/l^2$.
Thus, the Bekenstein-Hawking entropy and the Hawking temperature reads
\begin{equation}
    S_0=\frac{2\pi r_+}{4G_3}
\end{equation}
\begin{equation}
    T_H=\frac{r_+}{2\pi l^2}=\frac{G_3}{\pi^2 l^2}\hspace{1mm} S_0
\end{equation}
The first law of black hole thermodynamics which reads $dM=T_HdS_0$ gives
\begin{equation}
    C=\frac{dM}{dT_H}\sim S_0
\end{equation}
and 
\begin{equation}
    E=M\sim S_0^2
\end{equation}

Putting these expressions in Eq.(\ref{relmc}), we get (in the high-temperature limit and ignoring constants)
\begin{equation}
    \mathcal{S}\approx S_0-\frac{3}{2}\ln S_0-\ln S_0=S_0-\frac{5}{2}\ln S_0\label{btz}
\end{equation}
Now, in the case of both energy and volume fluctuations, $D$ and $V$ reads
\begin{align}
    D\sim P^2S_0^5\\
    V\sim S_0^2
\end{align}
Therefore, the microcanonical entropy for the case of simultaneous fluctuation in energy and volume is given by\cite{ghosh2022novel}
\begin{equation}
    \mathcal{S}=S_0-\frac{5}{2}\ln S_0
\end{equation}
which in the relativistic case in the high-temperature limit becomes
\begin{equation}
    \mathcal{S}=S_0-\frac{9}{2}\ln S_0
\end{equation}
Thus, we see that the contribution of the relativistic term is pronounced in the high-temperature limit which essentially signifies that the relativistic corrections indeed become appreciable in the high-temperature limit of a black hole and therefore must be included in the corrected microcanonical entropy. It is interesting to note that the second term of Eq.(\ref{btz}) is the same as the one found for the case of a non-rotating BTZ black hole when both energy and volume fluctuations are considered simultaneously. Furthermore, close to extremality, $T_H\approx0$, and the Eq.(\ref{relmc}) cannot be applied and the analysis breaks down. Therefore, we restrict ourselves to the case that is far from extremality such that the equation holds true.
\subsection{AdS-Schwarzschild Black Hole}
For a d-dimensional Schwarzschild black hole, $C=-(d-2)S_0$ and the Eq.(\ref{relmc}) cannot be applied signaling instabilities. Let us, therefore, focus our attention on the $d$-dimensional AdS-Schwarzschild black hole for which the metric reads
\begin{equation}
    ds^2=-\bigg(1-\frac{16\pi G_dM}{(d-2)\Omega_{d-2}r^{d-3}}+\frac{r^2}{l^2}\bigg)dt^2+\bigg(1-\frac{16\pi G_dM}{(d-2)\Omega_{d-2}r^{d-3}}+\frac{r^2}{l^2}\bigg)^{-1}dr^2+r^2d\Omega^2_{d-2}
\end{equation}
where $G_d$=$d$-dimensional Newton's constant, $d\Omega^2_{d-2}$ is the metric on unit $S^{d-2}$ and $\Omega_{d-2}$ is the area of this unit sphere. The
temperature and entropy reads (in the high-temperature limit)
\begin{equation}
   T_H\sim S_0^{1/(d-2)} 
\end{equation}
\begin{equation}
    C\sim S_0
\end{equation}
\begin{equation}
   E=M\sim S^{(d-1)/(d-2)}
\end{equation}
Thus, we obtain the microcanonical entropy in the high-temperature limit (ignoring constants)
\begin{equation}
    S\approx S_0-\frac{d}{2(d-2)}\ln S_0-\ln S_0
\end{equation}
The third term in the above expression is the relativistically corrected term and interestingly it does not depend on the dimension $d$ of AdS-Schwarzschild 
black hole. For the case of $4D$ AdS-Schwarzschild black hole when both energy and volume fluctuations are taken into account with $P=\frac{3}{8\pi l^2}$, the quantities $D$, $T$, $E$, and $V$ reads (in the high-temperature limit)
\begin{align}
    D\sim P^2S_0^3\\
    T\sim PS_0^{1/2}\\
    V\sim S_0^{3/2}\\
    E=M\sim S_0^{3/2}
\end{align}
Therefore, the corrected microcanonical entropy is given as\cite{ghosh2023statistical}
\begin{equation}
    \mathcal{S}=S_0-\frac{3}{2}\ln S_0
\end{equation}
while the relativistically corrected microcanonical entropy is obtained as
\begin{equation}
    \mathcal{S}=S_0-\frac{7}{2}\ln S_0
\end{equation}
So, for the case of $4D$ AdS-Schwarzschild black hole when both energy and volume fluctuations are considered, the relativistic correction is twice ($2\ln S_0$) that of the case when only energy fluctuations are considered.
\section{Conclusions}
In this paper, we studied relativistic correction to black hole entropy using the $\kappa$-generalized statistics which is a relativistic generalization of the Boltzmann-Gibbs statistics. We found that the relativistic corrections are logarithmic in nature and the equation of density of states gets additional multiplicative factors than the standard case which can be interpreted as Lorentz-like factors for the statistical ensemble under consideration. These relativistic corrections become appreciable only in the high-temperature limit suggesting that while taking this limit, the relativistic corrections must be included to get the correct entropy correction. It seems straightforward to extend these corrections to the three variables case such as the "open" ensemble\cite{ghosh2022logarithmic} following the method discussed in the paper and is therefore not considered here. Thus, formally speaking, the logarithmic corrections in the BG statistics is equivalent to the $\kappa\to0$ limit of the Kaniadakis statistics. Therefore, it seems more natural to work in this generalized statistics than the BG statistics. 
\appendix
\section{Kaniadakis statistics($\kappa$-statistics): A Brief Review}
\numberwithin{equation}{section}
$\kappa$-statistics is a relativistic generalization of the Boltzmann-Gibbs(BG) statistics. The $\kappa$-entropy emerges from the relativistic generalization of the Boltzmann-Gibbs-Shannon(BGS) entropy and generates power law-tailed distribution which in the limit $\kappa\to0$ reproduces the ordinary exponential distribution. This $\kappa$-generalized statistics has been applied successfully to a wide range of problems. Formally, it is a one-parameter deformation of the ordinary exponential and logarithmic functions as follows
\begin{equation}
    e_\kappa(x)=(\sqrt{1+\kappa^2x^2}+\kappa x)^{\frac{1}{\kappa}}
\end{equation}
\begin{equation}
    \ln_\kappa(x)=\frac{x^\kappa-x^{-\kappa}}{2\kappa}
\end{equation}
The $\kappa$-exponential and $\kappa$-logarithm for the case $0<\kappa<1$ can also be written as
\begin{equation}
    e_\kappa(x)=e\bigg(\frac{1}{\kappa}\arcsinh(\kappa x)\bigg)\label{ekappa}
\end{equation}
\begin{equation}
    \ln_\kappa(x)=\frac{1}{\kappa}\sinh\bigg(\kappa(\ln(x))\bigg)
\end{equation}
Some of the basic properties of the $\kappa$-exponential are as follows:
\begin{align}
    e_\kappa(x)\in\mathbb{C}^\infty(\mathbb{R})\\
    \frac{d}{dx}e_\kappa(x)>0\\
    \frac{d^2}{dx^2}e_\kappa(x)>0\\
    e_\kappa(-\infty)=0^+\\
    e_\kappa(+\infty)=+\infty\\
    e_\kappa(0)=1\\
    e_\kappa(-x)e_\kappa(x)=1
\end{align}
For a real number $r$, the following property holds
\begin{equation}
    [e_\kappa(x)]^r=e_{\kappa/r}(rx)
\end{equation}
Similarly, the $\kappa$-logarithm has following basic properties:
\begin{align}
  \myln(x)\in\mathbb{C}^\infty(\mathbb{R^+})\\
  \frac{d}{dx}\myln(x)>0\\
  \frac{d^2}{dx^2}\myln(x)<0\\
  \myln(0^+)=\infty\\
  \myln(1)=0\\
  \myln(+\infty)=+\infty\\
  \myln(1/x)=-\myln(x)
\end{align}
For a real number $r$, the following property holds
\begin{equation}
    \myln(x^r)=r\ln_{r\kappa}(x)
\end{equation}
For any $x$, $y$$\in\mathbb{R}$ and $|\kappa|<1$, the $\kappa$-sum is defined as
\begin{equation}
    x\myoplus y=x\sqrt{1+\kappa^2y^2}+y\sqrt{1+\kappa^2x^2}
\end{equation}
which is equivalent to
\begin{equation}
    x\myoplus y=\frac{1}{\kappa}\sinh(\arcsinh(\kappa x)+\arcsinh(\kappa y))
\end{equation}
The two important relations based on $\kappa$-sum, useful for our discussion are:
\begin{align}
  e_\kappa(x\myoplus y)=e_\kappa(x)e_\kappa(y)\\
  \ln_\kappa(xy)=\ln_\kappa(x)\myoplus\ln_\kappa(y)
\end{align}
Finally, we define $\kappa$-Laplace transform and its inverse as
\begin{align}
    F_\kappa(s)=\mathcal{L}_\kappa\{f(t)\}(s)=\int_0^\infty f(t)[\mye(-t)]^s dt\\
    f(t)=\mathcal{L}_\kappa^{-1}\{F_\kappa(s)\}(t)=\frac{1}{2\pi i}\int_{c-i\infty}^{c+i\infty} \frac{F_\kappa(s)[\mye(t)]^s}{\sqrt{1+\kappa^2t^2}}ds\label{lapinv}
\end{align}
The ordinary Laplace transform and its inverse are recovered in the limit $\kappa\to0$.
\section{Relation between $\kappa$-deformed entropy and $\kappa$-deformed partition function}
We start with the $\kappa$-distribution given as
\begin{equation}
\rho^i_\kappa=\frac{e_\kappa(-\beta E_i)}{\mathcal{Z}_\kappa}
\end{equation}
$\mathcal{Z_\kappa}$ is the normalization constant called the $\kappa$-deformed partition function. The $\kappa$-deformed entropy gives
\begin{align}
    S_\kappa=-\sum_i\rho^i_\kappa\ln_\kappa \rho^i_\kappa=-\sum_i \rho^i_\kappa [\ln_\kappa(e_\kappa(-\beta E_i))\myoplus(-\ln_\kappa\mathcal{Z}_\kappa)]\\
    =-\sum_i \rho^i_\kappa[-\beta E_i\myoplus(-\ln_\kappa\mathcal{Z}_\kappa)]
\end{align}
Here, we have used the properties of $\kappa$-deformed log, $\ln_\kappa (xy)=\ln_\kappa x\myoplus\ln_\kappa y$ and $\ln_\kappa(\frac{1}{x})=-\ln_\kappa x$. Using $\kappa$-sum, we obtain the above relation as
\begin{equation}
    S_\kappa=\sum_i \rho^i_\kappa[\beta E_i\sqrt{1+(\kappa\ln_\kappa\mathcal{Z}_\kappa})^2+\ln_\kappa\mathcal{Z}_\kappa\sqrt{1+(\kappa\beta E_i)^2}]
\end{equation}
Since $\sum_i\rho^i_\kappa=1$ and $\sum_i\rho^i_\kappa E_i=\langle E\rangle=U$, we have
\begin{align}
    S_\kappa=\beta\bigg(\sum_i\rho^i_\kappa E_i\bigg)\sqrt{1+(\kappa\ln_\kappa\mathcal{Z}_\kappa)^2}+\ln_\kappa\mathcal{Z}_\kappa\sqrt{\bigg(\sum_i\rho^i_\kappa\bigg)^2+\bigg(\kappa\beta\sum_i\rho^i_\kappa E_i\bigg)^2}\\
    =\beta U\sqrt{1+(\kappa\ln_\kappa\mathcal{Z}_\kappa)^2}+\ln_\kappa\mathcal{Z}_\kappa\sqrt{1+(\kappa\beta U)^2}
\end{align}
This gives
\begin{equation}
    S_\kappa=\ln_\kappa\mathcal{Z}_\kappa\myoplus\beta U
\end{equation}
which is equal to Eq.(\ref{kentropy2}) of the main text.
\section*{Conflict of Interest}
The author declares no conflict of interest.
\section*{Data Availability}
Data sharing is not applicable to this article as no datasets were generated or analyzed during the current study.

\bibliography{bib}
\bibliographystyle{unsrt}

\end{document}